\documentclass[aps,pre,showpacs,floats,twocolumn,amsfonts,amssymb,unsortedaddress,floatfix]{revtex4}
\input epsf
\usepackage{bm}
\usepackage{amsmath}
\usepackage{graphicx}

\begin{document}

\title{Quasi-One-Dimensional Thermal Breakage}
	
\author{Cristiano~Nisoli$^{1}$, Douglas Abraham$^{1,2}$, Turab Lookman$^{1}$ and Avadh Saxena$^{1}$}
\affiliation{$^{1}$\mbox{Theoretical Division and Center for Nonlinear Studies, Los Alamos National Laboratory, Los Alamos, NM, 87545 USA} \\
$^{2}$Rudolf Peierls Centre for Theoretical Physics, 1 Keble Road Oxford, OX1 3NP England}

\date{\today}
\begin{abstract}
Breakage  is generally understood in mechanical terms, yet  nano-structures  can  rupture not only under external loads but also via thermal activation. Here we treat in a general framework the statistical mechanics of thermally induced breakage at the nano-scale for one-dimensional systems. We test it on a simple approximation and find that  the probability of breakage controls  distinct regimes,  characterized by  sharp  crossovers and narrow peaks in the thermal fluctuations and specific heat. Our work provides predictions on clustering of new phases, of relevance in nano-fabrication. 
\end{abstract}

% nanoscale structures 

\maketitle

\section{Introduction}

Mesoscopic quasi-one-dimensional structures (Q1DS), such as nano-wires~\cite{Wang, Hernandez, Fan},  polymers~\cite{poly, DeGennes}, and biological molecules~\cite{prot} are of great current interest in science and technology. Such systems can now be manipulated at a molecular level and engineered to a desired specification using patterned substrates as templates~\cite{Neuman, Seeman, Wang, Hernandez, Fan, Adelung, Alaca}. 
As a result, there is a growing theoretical appreciation for subtle phenomena activated or suppressed by thermal fluctuations in {\it individual} systems: e.g. bubble opening and DNA denaturation under external load, nano-island to nano-wire transitions in Stranski-Krastanov growth, local narrowing in nano-wires, local collapse in nano-tubes, selection of growth direction in vapor deposition, or nano-dot to nano-ring transition in droplet epitaxy~\cite{Bishop, NisoliDNA, Nisoli2, Nisoli, Hernandez, Alaca, Adelung, Fan,Tersoff, Cao, Cao2, Chen, Zhu,Zhou}. 

In particular,  deposition on patterned substrates~\cite{Hernandez, Fan, Alaca, Adelung} could provide a method to  fabricate large networks of nano-wires  in arbitrary  combinations, and therefore to deposit nano-circuitry. These methods imply horizontal growth into patterned nano-channels obtained by strain-controlled cracking of the substrate~\cite{Alaca, Adelung}, or other means~\cite{Fan}.

However, no general framework exists to treat  {\it thermally induced full breakage} in Q1DS, despite  its emergence in supramolecular chemistry of non-covalent polymers~\cite{Sijbesma,deGreef} or of vapor deposition of nano-wires~\cite{Wang, Chen, Zhu, Zhou}. 
Theoretically, this problem  is quite  distinct  from the study of bistabilities,  instabilities or local failures~\cite{Langer, Nisoli2} where narrowing, collapsing or kinking, but not breakage,  produces a local change in functionality. In this article we introduce a statistical mechanical analysis of the thermally induced breakage based on Fisher's renewal equation approach~\cite{Fisher}.

The paper is structured as follows: in section II we describe the general approach. In section III we derive the double renewal equation for the problem of mechanical breakage in a Q1DS. In section IV, we show how to calculate the partition function $\Omega_n$
for an unbroken nano-wire in a channel in a simple approximation. We also explain criteria for forming
such wires with a defined width independent of the channel width. This reduces to an equivalent 1-d
Ising problem. We discuss the implications of this in some detail in section V. Although, as is well known, there are no phase transitions in such systems, in scaling theory, there are cross-over regions which we describe. These may well prove relevant in experimental situations.  

\section{Approach}

We study complete breakage by treating distinct, broken fragments. To set the scene, let us consider a sub-critical {\it planar} Ising ferromagnet. The magnetized domains are typically rectangles with rounded corners, one dimension of which will typically be reduced by imposing strip boundary conditions. 

Entropic repulsion~\cite{Fisher} from the lattice edges will compress domains laterally, tending to elongate domains, not necessarily to form wires. Yet, entropic repulsion between the sides of the domain on average parallel to the strip extrema will promote widening of the putative nano-wire segment. 

We suspect strongly that these entropic interactions will not suffice to produce the desired narrow, wire-like structures with a width asymptotically independent of the channel width. In support of this, we have available exact calculations on the 2-d
Ising ferromagnet in a strip~\cite{DBA} which demonstrate clearly that a nano-wire with
width defined by the interaction but not the channel width is not formed. 

There are a number of possible ways of achieving this. For instance, introducing 4-body interactions in a model of Baxter type can induce a binding-unbinding transition; in the bound state, the nano-wire segment has a width essentially independent of the strip width provided this is wide enough--yet  not  too wide, otherwise a Privman-Fisher argument~\cite{Privman} indicates that nano-segments will multiply laterally and thus our model is no longer applicable. 

Other factors with potential to control the width are differential chemical potential and elastic strain. A further crucial problem in discussing nano-wires is that of analyzing configurations in the channel (see Fig. 1) for connectedness at the nano-level. A development of the Fisher approach to surface phase transitions through the theory of recurrent events on the line~\cite{Fisher} is ideally suited to this task, as we shall show.

As a special case of this formalism, we establish a connection with the 1-d Ising model on a line with special boundary conditions which is useful for discussing our results. We emphasize that our model is a great deal more general and that the Ising model of itself does not contain our results. The only additional technique needed is the quantum mechanics of a particle on a finite line with a localized potential.

We establish and discuss conditions for the formation of nano-wires via epitaxial growth on a pre-patterned substrate, in a channel. A key point  is that given the channel width  the temperature should be necessarily low in a controlled way for nano-wires to form.  Further, we  show that when breakages coalesce in  Q1DS they  give rise to  narrow crossovers: steep changes of relevant observables and  narrow peaks in thermal fluctuations. These crossovers might have escaped the attention of experimentalists as pathological to desired result, e.g. the realization of long nano-wires in vapor deposition. Yet they 
  are indicative of new phases of clustering which could  be explored for nano-manufacturing. 

In~[\onlinecite{Nisoli, Nisoli2}], we considered the malfunction of a Q1DS (a nano-wire, a nano-tube or a bio-polymer) that comes about because two edges of the wire approach too closely as a result of thermal fluctuations. This model was treated using the Fisher recurrence idea: with the partition function $Z_n$ given by, 
\begin{equation}
Z_n=\Omega_n+\sum_{m=1}^{n-1}\Omega_m Z_{n-m}.
\label{old}
\end{equation}
This equation states that any configuration either has no malfunction (the first term on the right
hand side) or a first malfunction at position $m$, followed by a configuration with no restriction. Here $\Omega_n$,  is the partition function for the system of length $n$ with no points of malfunction. Equation (\ref{old}) can be converted into an algebraic one between generating functions by noting that the generating function of the convolution on the right hand is a product of the generating functions of  $\Omega_n$ and $Z_n$.

In this work, we consider {\it complete} breaking of nano-wires into separate parts. Unlike the macro-domain, where we would be discussing mechanical rupture, we have to consider once again the effects of thermal fluctuations. To construct a sensible model, we have to place the putative nano-wire forming material in a channel of finite width, otherwise the components will be found asymptotically far apart at equilibrium.

\section{Formalism}

Our approach is general, but for definiteness, we employ the language of epitaxial growth.  
Consider a straight channel of lateral size $D$ and length $Na$ ($a$ is a suitable discretization length and $N$ a positive integer; we take $a=1$) in which a nano-wire is grown in equilibrium with gas as in Fig.~1. 
A piece of the system of length $na$ might contain a mix of gas and (possibly) broken nano-wires--and then we call ${\cal Z}_n$ its 
partition function; or gas and a single unbroken nano-wire---${\cal O}_n$; or only gas---${\cal G}_n$.  Both ${\cal O}_n$  and ${\cal G}_n$   are known while ${\cal Z}_n$ is to be found.

The system can: (I) contain a single unbroken nano-wire, contributing ${\cal O}_N$ to  ${\cal Z}_N$; (II) contain a shorter nano-wire of length $j$ followed by gas,  contributing ${\cal O}_j \times {\cal G}_{N-j}$ to ${\cal Z}_N$; (III) or  can start with a nano-wire of length $i$, be followed by a portion of  gas of length $j$, and then whatever else is allowed on the remaining length $N-i-j$, contributing  ${\cal O}_{i} \times {\cal G}_j \times {\cal Z}_{N-i-j}$ to ${\cal Z}_N$.

By summing over all these configurations, one obtains the following double convolution equation 
\begin{eqnarray}
{\cal Z}_N&=&{\cal O}_N \nonumber \\
&+& {\cal O}_{N-1}{\cal G}_1+{\cal O}_{N-2} {\cal G}_2+ \dots +{\cal O}_1{\cal G}_{N-1} \nonumber \\
&+&{\cal O}_{N-2} ({\cal G}_1 {\cal Z}_1)\nonumber \\
&+&{\cal O}_{N-3} ({\cal G}_1 {\cal Z}_2+{\cal G}_2 {\cal Z}_1) \nonumber \\
&+&\dots  \nonumber \\
&+& {\cal O}_1\left({\cal G}_{N-2} {\cal Z}_1+{\cal G}_{N-3} {\cal Z}_2+\dots+{\cal G}_1 {\cal Z}_{N-2}\right) , 
\label{conv}
\end{eqnarray}
($N\ge1$ and ${\cal Z}_n$, ${\cal O}_n$, ${\cal G}_n$ are taken to be  zero if $n\le 0$.) 

\begin{figure}[t!]   
\includegraphics[width=3. in]{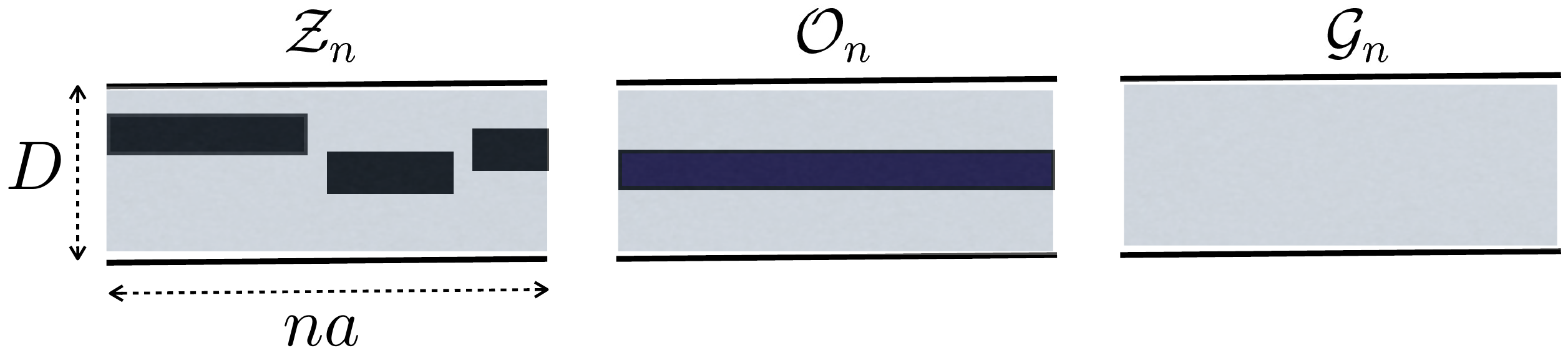}
\caption{Schematics of a nano-wire  growing in a channel of lateral size $D$, in equilibrium with its gas. ${\cal Z}_n$ is the partition function (unknown) for  a mix of gas and broken nano-wires, of length $na$.  ${\cal O}_n$ reppresents the (known) partition function for a portion of the channel containing a single unbroken wire and gas. ${\cal G}_n$ is the (known) partition function for a portion of the channel containing only gas. (The wires can  bend or grow in different shapes, we draw them straight for simplicity.)}
\label{drawing}
\end{figure}

The linear density of free energy for the entire system is then ${\cal F}= -\beta^{-1} \ln ({\cal Z}_N)/N$ in the  limit of large $N$, and as usual $\beta^{-1}\equiv k_B T$. 

Since the gas phase approaches the thermodynamic limit faster than the solid phase, we take ${\cal G}_n$ to be exponential in $n$. 
Then (\ref{conv}) maintains the same form in the {\it relative} (to the gas phase)  quantities $Z_n={\cal Z}_n/{\cal G}_n$, $\Omega_n={\cal O}_n/{\cal G}_n$, $V_n={\cal G}_n/{\cal G}_n=1$. The relative  free energy is thus   $\phi= -\beta^{-1} \ln (Z_N)/N$. 

We solve (\ref{conv}) by noting its convolution structure
\begin{equation}
Z_N=\Omega_N+(\Omega \ast V)_N+(\Omega * V * Z)_N
\label{conv2}
\end{equation}
and introducing  the generating functions
\begin{eqnarray}
\tilde Z (u)&=&\sum_{n=1}^{\infty} Z_n u^n,~\tilde \Omega (u)=\sum_{n=1}^{\infty} \Omega_n u^n, \nonumber \\
\tilde V(u)&=&\sum_{n=1}^{\infty} V_n u^n={u}/(1-u).
\end{eqnarray}

Since the convolution product is associative, the generating function of the third term on the right side of ~(\ref{conv2}) is an algebraic product of generating functions. Since $\Omega_n$, and therefore its generating function $\tilde\Omega$   are assumed known, $\tilde Z$ can be obtained as:
\begin{equation}
\tilde Z(u)%=\frac{\tilde \Omega(u)(1+\tilde V(u))}{1-\tilde \Omega(u) \tilde V(u)}
={\tilde \Omega(u)}/{\left[1-(\tilde \Omega(u) +1)u\right]}.
\label{Zu}
\end{equation}
Then the partition function $Z_n$  is obtained from  (\ref{Zu}) by contour integration and the residue theorem. For this to be useful, we need the analytic structure of $\tilde \Omega(u)$. 

\section{General Case}

Thus far, we have extended the renewal ideas of Fisher~\cite{Fisher} to tackle the
problem of connectedness of nano-wire configurations. %We have illustrated this with a simple assumed form for $\Omega_n$ that surely does not exhaust the complexity of real possibilities. 
We need now to analyze $\Omega_n$  in a critical way, and provide a rationale for the applicability and generality of the previous analysis, at least in the case of vapor deposition of nano wires. 

A putative nano-wire inserted into a strip of ``gas'' is a strip of solid inserted between two boundaries running roughly parallel to the strip axis. Each boundary has an independent incremental free energy, but there may be additional interactions due to differential fugacity, elastic mis-match and 4-body terms, to name but three. These can produce bound states between the opposite sides of the wire, with the nano-wire width determined by these parameters,  considerably less than, and only weakly dependent on, the channel width $D$ (we measure  lengths in units of $a$ and $D$ is a pure number). 

The key idea is that one side of the nano-wire can partially wet the other side. Thus we are in the familiar wetting, or equivalently pinning-depinning territory~\cite{Abraham1, Abraham2, Abraham3, Abraham4, Abraham}.
\begin{figure}[t!]   
\begin{center}
\includegraphics[width=2.9 in]{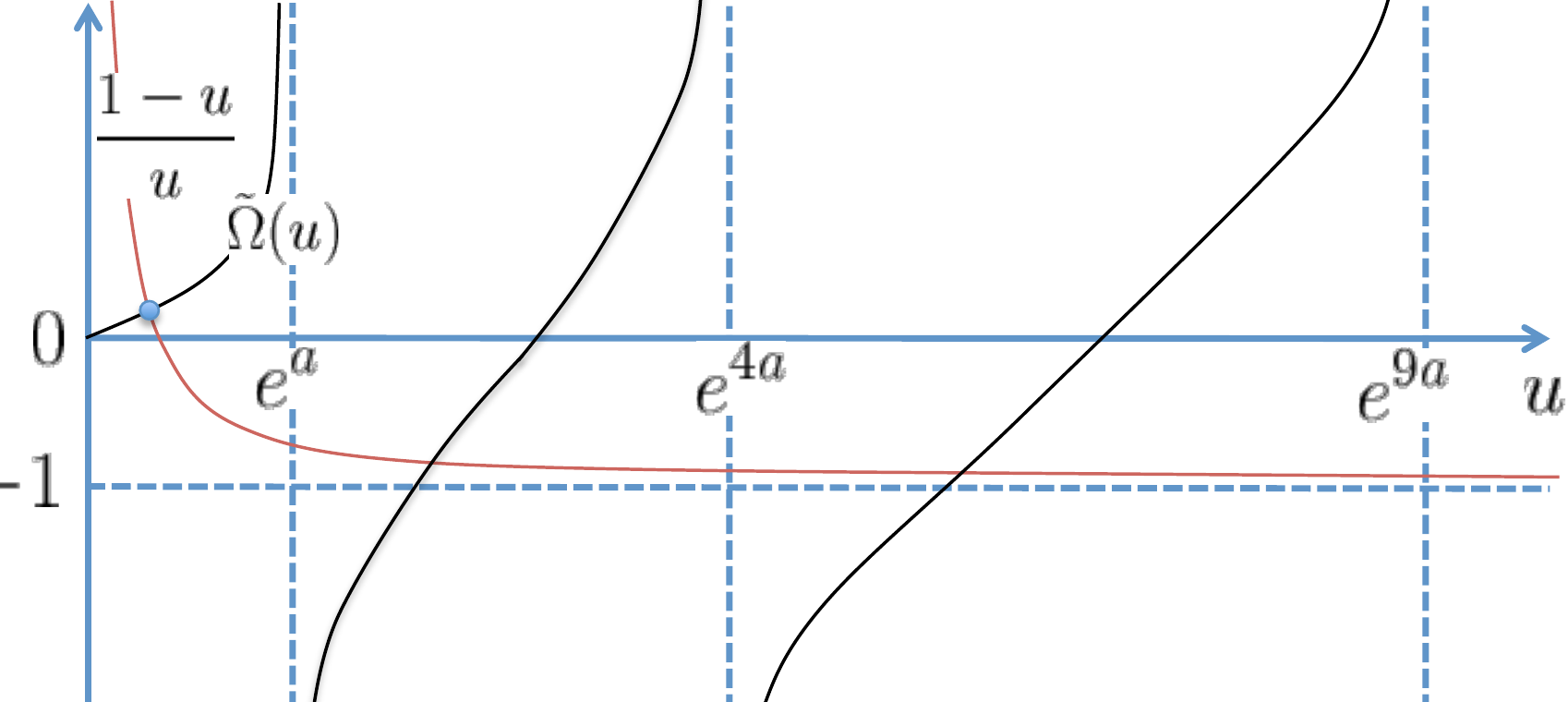}\vspace{-3mm}
\caption{(Color Online) Graphical solution of $\tilde \Omega(u)=1/u-1$:  The poles of $Z(u)$ are the intersections between the function $1-1/u$ (red line--light gray) and $\Omega(u)$ (solid black lines). The lowest root (blue circle) defines the free energy.}\vspace{0mm}
\end{center}
\label{zerosfig}
\end{figure}
Under these conditions, it makes  sense to treat the relative and center of mass coordinates independently. To a good approximation, the center of mass mode has an energy per unit length given by $E=\chi k^2$  and  takes discrete values given by the solution of an equation of the form $\exp(iDk)=f(k)$, where $2\chi$ is the interfacial stiffness~\cite{MPAFisher}, and $f$ is an analytic function~\cite{Abraham}. For instance, the transfer matrix solution of the planar Ising strip affords us just such an example, the energy expression being well-approximated by the free particle form above and the momenta being given to good approximation by $k_j=\pi j/D$, $j=1,...,D$.

In this  approximation we can show that:
\begin{equation}
\tilde\Omega(u)=p^2\sum_{j=1}^D q_j \omega u/\{ \exp\left[\chi \left(\pi j/D\right)^2\right]-\omega u\},
\label{OmegaD}
\end{equation}
with $q_ j = 2D^{-1} \cot^2 (\pi j / 2D)$. This has the interesting large $D$ behavior $q_j=2D/\pi$, a manifestation of {\it transverse} entropy. 

To complete the job, we have to solve for the zeros of the denominator in $\tilde Z(u)$, either by sketching a graph (Fig. 2) or otherwise. 
The key point is that there is a unique minimal such value, denoted $u_0$, which is the only one to report in the incremental free energy $\phi$. This justifies the interpretation of the ÒsolidÓ as a nano-wire and also
the ultimate application to this system of the Ò1-d IsingÓ analysis conduced below in~(\ref{ising}); the other,
non-minimal $u$ values are indeed there, but they do not report in the limiting free
energy $\phi$, as we have demonstrated.

 The formalism of this section would also enable us to analyze the fluctuations in the structure of individual nano-wires both as a function of the interaction parameters and of the nano-wire length, but this issue seemed to us to depart sufficiently from the theme of this work to make a separate treatment advisable.

\section{Case of No Finite-Size Effects}

To gain some insight into the structure of $\Omega_n$ we note that in general, and in the language of epitaxial growth, the free energy density of the solid phase $f$ is given by $\lim_{n\to \infty} n^{-1}\ln \Omega_n=\beta f$. 

\begin{figure}[t!]   
\begin{center}
\includegraphics[width=3.2 in]{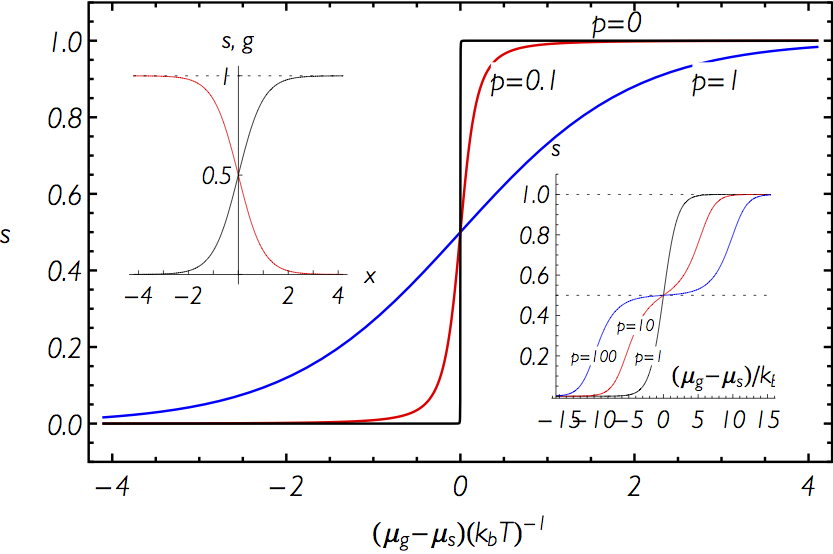}
\caption{(Color Online) Relative concentration of solid phase $s$ as a function of $(\mu_s-\mu_g)/k_B T$  for small  $p$, and for large $p$ (right inset): as $p$ grows  three distinctive phases appear; left inset: linear density of gas (light grey, red online) and solid (black) phase as a function of the parameter $x$.}
\end{center}
\label{s}
\end{figure}

For finite $n$ one can write $\Omega_n \propto  p^2_n \omega^n$ with $\lim_{n \to \infty}n^{-1}\ln p^2_n=0$. Clearly $\omega$ is the fugacity per unit length of the nano-wire while the series $p_n$ reflects finite size effects, possibly induced by elastic fields or other phenomena such as charge interactions, depending on the system at hand. 

The choice of a constant $p_n$  is instructive and more general than it appears, as we argued in the previous section. Consider then:
\begin{eqnarray}
\Omega_n=p^2 \omega^{n}
\label{basic0}
\end{eqnarray}
and thus
\begin{eqnarray}
\tilde\Omega(u)=p^2\omega u/(1-\omega u).
\label{basic}
\end{eqnarray}
Here again $\omega$ is the fugacity per unit length of the nano-wire and $p$ is the Boltzmann factor for a break, which does not depend on the length of the wire; (\ref{basic0}) is clearly an approximation describing systems in which finite size effects of the solid phase are negligible. 

To situate this approximation in the broader context described above we can compare (\ref{OmegaD})  with (\ref{basic}). We see that for the channel problem we had D terms.  Provided $\chi (\pi/D)^2\gg1$, in looking for the minimal solution of $\tilde\Omega(u)=u^{-1}-1$, we can treat terms with $j>1$  as a perturbation of the $j=1$  solution which we have already found; all that is needed is to replace $\omega$ and $p$ by
\begin{equation}
 \omega'=\omega \exp\left[-\chi (\pi/D)^2\right], ~~ p'=q_1^{1/2} p.
 \end{equation}
 Then to first order we can take over the calculation based on (\ref{basic}). The condition for this is, as stated above, that $D$ is small enough, or, given $D$, that the stiffness is large enough. This implies that the nano-wire will be relatively free of kinks, clearly a desideratum. Provided we can also satisfy  $2D\gg \pi$, then $q_1\sim 8D/\pi^2$. The linearity in width when the nano-wire is reasonably stiff makes good sense as a manifestation of transverse translational degeneracy. 

We can now turn to the solution of the problem for the special choice of $\Omega_n$ given by (\ref{basic0}). The zeros in the denominator of (\ref{Zu}), which give simple poles of the integrand, are given by the solutions of  $\tilde\Omega(u)=u^{-1}-1$, which in the case of (\ref{basic}) is simply a quadratic equation.
This model is a somewhat disguised version of a 1-d Ising chain with fixed spins on the extremities. Each segment of the nano-wire is represented by an up spin, whereas the gas has down spins. 

The energy of any configuration in the channel, with the same edge conditions as before, is:
\begin{equation}
\beta E=-2K \sum_{n=1}^{N-1}(1-\sigma_n \sigma_{n+1})/2+2h\sum_{n=1}^{N}(1+\sigma_n)/2
\label{ising}
\end{equation}
with $p=\exp(-2K)$ and $\omega=\exp 2h$. At the left hand side we always have an element of nano-wire, so that the extreme spin is up, but the spin at the other end is down. Thus identical answers can be obtained in either way:
\begin{equation}
Z_n=[\omega p^2/(\omega_+-\omega_-)]\left(\omega_+^n-\omega_-^n\right),
\label{Zp}
\end{equation}
where the    $\omega_{\pm}$, which are closely related to the $d=1$ Ising transfer matrix eigenvalues, are given by
\begin{equation}
2\omega_{\pm}=1+\omega \pm\left[{\left(\omega-1\right)^2 +4 \omega p^2}\right]^{1/2},
\label{omegap}
\end{equation}
with  $\omega_+>\omega_->0$. The limiting free energy $\phi$ per unit channel length is given by:  $\phi=-\beta \ln \omega_+$. 

\begin{figure}[t!]   
\begin{center}

\hspace{.7 mm}\includegraphics[width=3.28 in]{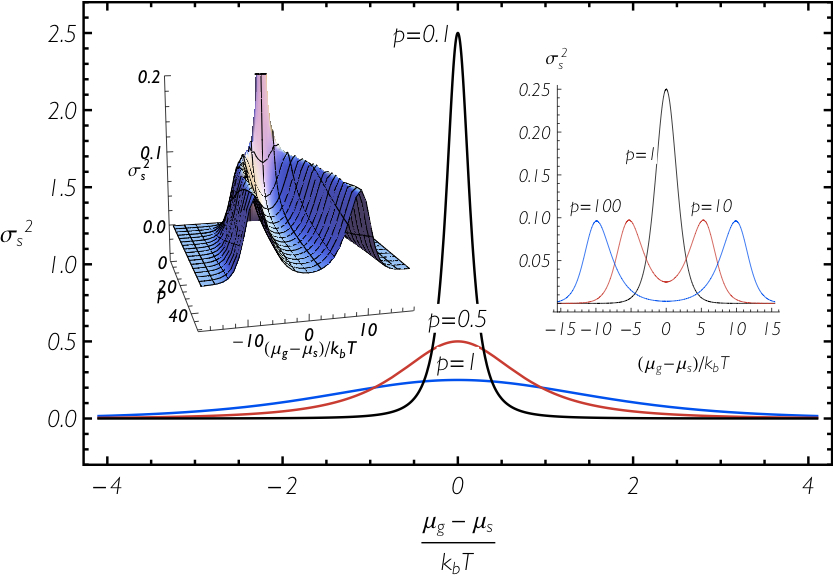}

\caption{Fluctuation of the concentration of solid phase, $\sigma^2_s$  vs.  $(\mu_s-\mu_g)/k_B T$: as $p\rightarrow0^+$ the peak around transition becomes sharper; for $p\gg1$ (right inset) it splits into two peaks  at $\mu_g=\mu_s\pm 2k_B T \ln(2p)$  separating the three phases (solid, gas and dense atomic clustering); left inset: the spliting of peaks as $p$ increases.}
\end{center}
\label{sigma}
\end{figure}

To study the ordering in this model, we can write $Z_N$  as a polynomial in $p$, the power being $n_b$, the number of breakages,  and the coefficient multiplying such a term independent of $p$. It is then easy to show that the fraction of breakages and its variance are given by:
\begin{equation}
n_b=-\beta p \partial_{p}\phi, ~\sigma^2_{n_b}=p\partial_p n_b.
\label{wnb}
\end{equation}
An analogous argument applies to the fraction $s$ of up spins or nano-wire, giving:
\begin{equation}
s= -\beta \omega \partial_{\omega}\phi,~ \sigma^2_s=\omega\partial_{\omega}s.
\label{wnb2}
\end{equation}

This gives directly:
\begin{equation}
s=\frac{\omega_+-1}{2\omega_+-\omega-1},~n_b=\frac{2(\omega_+-\omega)(\omega_+-1)}{(2\omega_+-\omega-1)\omega_+}.
\label{snb}
\end{equation}
The average channel length lacking a nano-wire is given by $g=1-s$, so the average length of a nano-wire is  $l_s= 2 s/ n_b$  and that of a gap is  $l_g=2 g/n_b$. Thus:
\begin{eqnarray}
l_s=\frac{\omega_+}{\omega_+-\omega}, ~ l_g =\frac{\omega_+}{\omega_+-1}.
\label{l2}
\end{eqnarray}
\begin{figure}[t!]   
\begin{center}
\includegraphics[width=3. in]{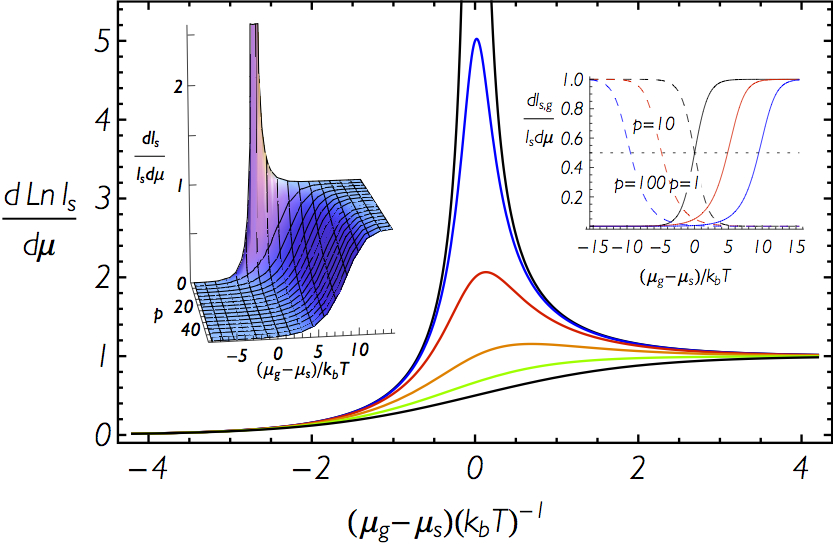}\vspace{-2 mm}
\caption{(Color Online) Logarithmic derivative of the average nano-wire length $l_s$ vs. $(\mu_s-\mu_g)/k_B T$ for small values of $p$: $p=0$ (black, diverging), $p=0.1$ (blue), $p=0.25$ (red), $p=0.5$ (orange), $p=0.75$ (green), $p=1$ (black). As $p\rightarrow0^+$ a peak grows corresponding  to the crossover from solid to gas, as   nano-clusters of length $1/p$ form, interspersed with intervals of gas of equal length $1/p$. Left inset: logarithmic derivative of  $l_s$ vs. $(\mu_s-\mu_g)/k_B T$ and $p$ showing the peak  at $p\ll1$ as well as a transition from dense atomic clustering region to long nano-wires, whose length grows exponentially with the  chemical potential.  Right inset: logarithmic derivative of $l_s$, $l_g$ vs. $(\mu_s-\mu_g)/k_B T$ for large values of $p$  [$p=1$ (black), $p=10$ (red), $p=100$ (blue)].} 
\end{center}
\label{dLs}
\end{figure}

We can check these equations for the obvious symmetry  $\omega \rightarrow 1/\omega$: then $\omega_+\rightarrow \omega_+/\omega$, $n_b \rightarrow n_b$,  $l_s\rightarrow l_g$. The condition for producing large $l_s$ is then $\omega_+-\omega \rightarrow 0+$, which is the case if $p\rightarrow 0$ and $\omega-1\rightarrow 0+$. In 1-d Ising language, this is completely obvious, as this is the critical point from one side. 

\section{Crossovers and Scaling Limits}

Note that if  $x$ is defined by $\sinh x=(\omega-1)/(2 p\sqrt{\omega})$, then 
\begin{equation}
s=\frac{\exp x }{2\cosh x}. 
\label{sx}
\end{equation}
Similarly, for the gas fraction, one has $g(x)=s(-x)$. Equation (\ref{sx}) provides the basis for the scaling limits at different crossovers. 

From~(\ref{sx}) we see that  for large positive values of $x$, $s$ tends to one and $g$ to zero, and conversely $s$ tends to zero and $g$ to 1 for large negative values of $x$ (Fig. 3, left inset). The right limit  corresponds to long wires separated by short ruptures of the size of a unit cell. Symmetrically, the left limit corresponds to long intervals of vapor interrupted by the occasional atomic-sized cluster of solid. Of course these limits do not correspond to the pure solid and pure gas phases, respectively:  for any nonzero value of $p$ there will always be breakage, no matter how diluted.

We can relate (\ref{basic0}) to thermodynamics via  $\omega=\exp(\beta \mu)$, $p= M \exp(-\beta \tau)$: $\mu=\mu_g-\mu_s$ is the difference in chemical potential between gas and solid per unit length, $M$ accounts for the lateral degeneracy of a broken piece, and $\tau$ is the energy cost of a rupture. 

Then $x$ generalizes the quantity $\beta \mu/2$ when $p\ne1$.  Indeed  $p\sinh x=\sinh\left(\beta \mu/2\right)$, and $x=\beta \mu /2$ when $p=1$.
 This corresponds to $K=0$ in the Ising model, or  noninteracting spins in a field, a pure paramagnetic phase, and could describe the case of a noncovalent polymer~\cite{deGreef,Sijbesma} inside a long nano-tube. For non-covalent polymers, $n$ in  our formalism  counts links rather than monomers: the energy of a
polymer of finite size $n$ is then $n \mu $, without any extra contribution
from a surface tension term. If the lateral size of the nanotube is then small
enough (not much larger than the diameter of the polymer), one can assume
no lateral degeneracy, and therefore p=1 seems to be  a good approximation.

\begin{figure}[t!]   
\begin{center}
\includegraphics[width=2.1 in]{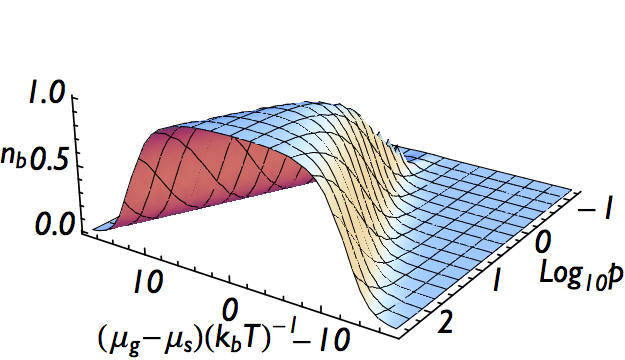}\\
\includegraphics[width=2.1 in]{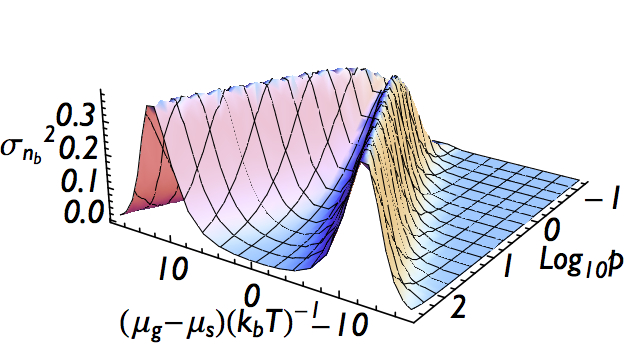}
\caption{Linear density of breakages $n_b$ (top) and its thermal fluctuations (bottom)  as a function of the breakage parameter $p$ and of $(\mu_s-\mu_g)/k_B T$.}
\end{center}
\label{nbfig}
\end{figure}

We now offer predictions in the context of nano-wire deposition, and in terms of the experimentally controllable parameters $\mu$ and $p$.  The new variable $x$ is defined only  when  $p\ne 0$. Yet  for $p=0$ both  $Z_n/p^2$ and $\Omega_n/p^2$ are still well posed in (\ref{basic}) and (\ref{Zp}). The assumpton $p=0$ simply means that breakages are forbidden, $p$ being the Boltzmann factor for a rupture. In the equivalent Ising model, it would imply an infinite coupling constant. 
 
 Clearly the case $p=0$  replicates  the bulk transition between solid and gas. In fact, when  $p=0$ and $\omega>1$ (i.e. $\mu_g>\mu_s$) from (\ref{omegap})  $\omega_+=\omega$, therefore from (\ref{snb}) $s=1$, $g=0$, $n_b$=0: the entire channel is filled with solid without any breakage.
Conversely when   $0<\omega<1$ (i.e. $\mu_g<\mu_s$) we have $\omega_+=1$, and thus $s=0$, $g=1$ and still $n_b=0$: the entire channel is filled with gas. The discontinuity at $\mu=0$ corresponds to a real transition from a solid-filled to gas-filled channel.   

While $p=0$ is unphysical, we can consider the {\it scaling} limit (denoted slim) $\omega\rightarrow1\pm$, $p\rightarrow 0$, such that $\sinh x $ has a definite limit, then: $\mathrm{slim} l_s=\mathrm{sech} x$ and $\mathrm{slim} l_g=\exp(-x)$. Thus, to have long nano-wires, take p small, then adjust the relative weights of wire and gas using $x$ as a control parameter. The behavior in this case is {\it ferromagnetic}. There is another scaling limit which has $|\ln \omega | \rightarrow\infty$, $p\rightarrow \infty$  such that $\sinh x=\mathrm{slim}[\pm \omega^{\pm1/2}/(2p)]$ (plus when $\omega\rightarrow\infty$ and minus when $\omega \rightarrow 0+$). This is {\it antiferromagnetic} in character. It corresponds to large lateral entropy.

Particularly interesting is the    ``ferromagnetic'' scaling limit $p\rightarrow 0^+$, $\omega\rightarrow1\pm$, which corresponds, as we will see, to a crossover through nano-clustering. Plots of the fraction of solid $s$ (Fig. 3) and  its fluctuations (Fig. 4) vs. the difference in chemical potential $\mu$ show a sharp crossover around $\mu=0$ in the limit of small $p$. As $p\rightarrow 0^+$, the crossover becomes sharper, as the plot for $s$ tends to a step, and a narrow peak in thermal fluctuations grows around $\mu=0$: the system tends asymptotically to the ``bulk transition'' described above  ($p=0$). 
For small $x$, one can  approximate $x\sim\beta \mu/p$ and thus the width of the  crossover in terms of  the chemical potential difference  is $\beta \Delta \mu \simeq 2p$.  Figure~5 shows the logarithmic derivative of the average length of the nano-wires. For small $p$, above (below) the crossover,  the average length of the wires (gaseous portions) grows exponentially in the difference of chemical potentials. 

At crossover  we have formations of equally spaced nano-clusters. Equation~(\ref{wnb}) shows that for $\mu=0$ one has 
\begin{equation}
n_b=p/(1+p)
\end{equation}
and therefore 
\begin{equation}
l_s=l_g=(1+p)/p\simeq1/p
\end{equation}
 for small $p$. 
 
 Since at $\mu=0$ one has $s=g=1/2$,  the crossover corresponds to nano-structures of finite length $a/p$  separated by intervals of equal length. This crossover through nanoclustering is signaled not only by peaks in fluctuations, but also the scaling limit mentioned above. This  implies, experimentally,  a {\it data collapse} in the scaling parameter 
\begin{equation}
y\equiv (\omega-1)/2p.
\end{equation}
Indeed the expression for $\omega_+$ in (\ref{omegap}) returns
\begin{equation}
\mathrm{slim} 2 y_{+} =y+\sqrt{y^2+1}, 
\label{yo}
\end{equation}
where $y_{+}\equiv (\omega_{+}-1)/2p$, and as above the scaling limit corresponds to $p$ and $(\omega-1)\rightarrow 0+$ with $y$ finite. It is then simple to prove that $\mathrm{slim}\sinh x = y$ and therefore $s$  is a function solely of the scaling parameter $y$. 

Since $\omega_+=1+p\exp x$, the linear density of free energy for the system ${\cal F}=-k_B T \ln (\omega_+)+\mu_g$ at first order in $p$ is ${\cal F}\simeq\mu_g -p k_B T  \exp x $; from (\ref{snb}) we  obtain  $n_b=p\mathrm{sech} x $ where $\sinh x=y$. Then the expansion of ${\cal F}$ at $x=0$ returns  ${\cal F}\simeq -k_B T p+(\mu_g+\mu_s)/2+(\mu_g-\mu_s)^2/(8 k_B T p)$: the first term is the  expression for the free energy of breakages between degenerate phases and can be obtained via heuristic treatments; the second term  reaffirms the equal mixture of solid and gas at crossover ($s=g=1/2$); the third term shows that the  lowest order contribution in the difference between chemical potentials is quadratic: this ensures a  peak in the specific heat, which diverges {as $p\rightarrow 0$}.
This nano-clustering crossover should be accessible when the surface energy cost of a breakage (which is proportional to the size of the channel or pore) is much larger than temperature.

 The antiferromagnetic limit is achieved at large $p$, which models growth in a broad yet  {\it not too broad} channel (when the lateral entropy and therefore $M$ are large, see above): a larger channel implies larger entropy gained by the broken pieces as they can move laterally. We show next that it corresponds to a phase of interspersed  atomic clustering.
Figure 5, right inset, shows formation of   plateaus in the ratio of gas and solid phase  at large $p$.  Figure~6, top panel, shows that, for large $p$, $n_b\simeq1$ and therefore, since $s=g=1/2$ on the plateau, then $l_s=l_g=1$. For large $p$, the system can access an antiferromagnetic phase of adsorbed solid and gas alternating at the unit cell length scale. We will show below that effective transitions from that region to the phases of $s=1$ (adsorbed solid) and of $s=0$ (gas) occur at  $\mu/k_B T=\pm 2\ln(2p)$. These crossovers are characterized by peaks in the fluctuation of both the solid fraction and of the density of breakages (Fig.~3 insets, Fig.~5, bottom panel).   

The physical picture is the following: as before, when $\mu/k_BT$ is large and positive, the system is made of long wires interrupted by sparse and short breakages. When  $\mu/k_BT$ drops to $2\ln(2p)$, the density of breakages suddenly increases (Fig. 6, top panel), as do fluctuations (Fig. 6, bottom panel), and the average length of the adsorbed wires $l_s$ shortens {\it while the average length of the gas portions  stays $l_g\sim 1$}--as shown by the absence of divergence in the logarithmic derivative of $l_s$ or $l_g$ at crossover (Fig. 5 insets). %, while $l_s$ decreases all the way to $a$. 
 At this point, the system enters the phase of interspersed  atomic clustering  until  $\mu/k_BT\rightarrow -2\ln(2p)$, at which point the clusters begin to separate in a phase of long gaseous gaps  separated by atomic  clusters.
 
Using the  scale limit argument one can show readily that data collapse depends now on the scaling parameter for the left and right crossover 
\begin{equation}
y^{r,l}_{\infty}\equiv\omega^{\pm 1/2}/(2p),
\end{equation}
for which $\sinh x= y^{r,l}$ depending on the side of the crossover. Since $y^{r,l}_{\infty}=\exp\{\left[\pm\beta \mu-2\ln (2p)\right]/2\}$,  the crossovers occur at $\mu/k_B T=\pm 2\ln(2p)$. These crossovers are characterized by peaks in the fluctuation of both the solid fraction and of the density of breakages (Fig.~3 and 5). 

For Stranski-Krastanov deposition~\cite{Stranski}, Tersoff and Tromp have predicted on purely energetic grounds~\cite{Tersoff}   the growth of wires of defined lateral size as the result of a shape-anisotropy transition. Elsewhere we had investigated the thermal stability of this shape anisotropy transition~\cite{Nisoli} to explain further experimental results~\cite{Zhu, Zhou}. The current analysis completes our previous treatment and shows that even at a temperature at which the shape anisotropy transition is stable, long nano-wire fabrication can be compromised by thermally activated  breakage.

\section{conclusion}

In summary, we have introduced an extension of Fisher's renewal theory of
surface phase transitions~\cite{Fisher} and analyzed the stability toward thermal breakage of Q1DS. We have employed the treatment to investigate the connectedness of nano-wires assembled
 in a channel when complete rupture can occur. 
 We have argued that in the case of breakage of nano-wire growth by vapor deposition on a pre-patterned substrate the formalism is reminiscent of an emergent Ising model.  We have shown that this brings about two crossovers corresponding to a ``ferromagnetic'' and an ``antiferromagnetic'' regime (in the Ising terminology). The first corresponds to a crossover through formation of equally-spaced nano-clusters, and it is achieved in regimes of high rupture cost. The opposite regime corresponds to a broad (in the difference of chemical potentials) phase of atomic clustering. We provide theoretical conditions for facilitating and controlling nano-wire growth, and to exploit thermal breakage in nano-structuring.

Our method has potential
for application in a much wider setting; all that is needed is to calculate $\Omega_n$. This can be computed in different ways for the problem at hand. When lateral fluctuations of the system are relevant, such as in the growth of nanowires on a vicinal surface~\cite{Yu}, $\Omega_n$ can be computed by summing over the lateral fluctuations in a path integral and then solving it as an equivalent Schr\"odinger problem~\cite{Nisoli, Nisoli2}. Also, finite size effects in $\omega_n$ due to interactions, elastic fields or charges, can bring true phase transitions.  We will report on this in future work.

This work was carried out under the auspices of the National
Nuclear Security Administration of the U.S. Department of
Energy at Los Alamos National Laboratory under Contract
No. DEAC52-06NA25396.

\end{document}